\documentclass[pre,aps,twocolumn,showpacs,floatfix,superscriptaddress]{revtex4-1}
\usepackage{graphicx}
\usepackage{bm}
\usepackage{amsmath}
\usepackage{amssymb}
\usepackage{amsfonts}
\usepackage{dcolumn}%
\usepackage{color}
\usepackage{float}
\usepackage{soul}

\begin{document}


\title{Dramatic effect of fluid chemistry on cornstarch suspensions: linking particle interactions to macroscopic rheology}

\author{Loreto Oyarte G\'alvez}
\affiliation{Physics of Fluids group, University of Twente, P.O. Box 217, NL-7500 AE Enschede, The Netherlands}
\author{Sissi de Beer}
\affiliation{Materials Science and Technology of Polymers, University of Twente, P.O. Box 217, NL-7500 AE Enschede, The Netherlands}
\author{Devaraj van der Meer}
\affiliation{Physics of Fluids group, University of Twente, P.O. Box 217, NL-7500 AE Enschede, The Netherlands}
\author{Adeline Pons}
\email[]{adeline.pons@normalesup.org}
\affiliation{Physics of Fluids group, University of Twente, P.O. Box 217, NL-7500 AE Enschede, The Netherlands}

\date{\today}
\begin{abstract}

Suspensions of cornstarch in water exhibit strong dynamic shear-thickening. We show that partly replacing water by ethanol strongly alters the suspension rheology.  
{We perform steady and non-steady rheology measurements combined with atomic force microscopy to investigate the role of fluid chemistry on the macroscopic rheology of the suspensions and its link with the interactions between cornstarch grains.} Upon increasing the ethanol content, the suspension goes through a yield-stress fluid state and ultimately becomes a shear-thinning fluid. On the cornstarch grain scale, atomic force microscopy measurements reveal the presence of polymers on the cornstarch surface, which exhibit a cosolvency effect. At intermediate ethanol content, a maximum of polymer solubility induces high microscopic adhesion which we relate to the macroscopic yield stress.

\end{abstract}
\date{\today}
\maketitle

Suspensions are mixtures {of} undissolved particles in a liquid. They are literally found all around us: 
mud, paints, pastes and blood \citep{Wagner_2009}. 
{T}he viscosity of a dense suspension can vary by orders of magnitude in a small shear rate interval \citep{Barnes_1989}. 
{S}ubjected to an increasing shear rate, dense suspensions first tend to become less viscous (shear-thinning) and then more viscous (shear-thickening). 
The viscosity of some suspensions, especially non-Brownian ones, {may} 
increase 
{so much} that they effectively become solid \citep{Roche_2013}. Although standard rheology measurements provide a great tool to study this phenomenon \citep[e.g.][]{Fall_2012,Brown_2010}, they are mainly limited to steady-state 
conditions.
Many studies point out that dense suspensions exhibit remarkable dynamic phenomena emerging under non-steady-shear conditions: stable holes in thin vibrated layers \citep{Merkt_2004}, non-monotonic settling  \citep{vonKann_2011}, dynamic compaction front \citep{Waitukaitis_2012} or fracturing \citep{Roche_2013}. 
Oscillatory rheology helps to describe some of these dynamic behaviours \citep{Deegan_2010}, but remains limited to constant volume conditions.

{D}ynamic shear-thickening has been widely investigated \citep{Brown_2014}, but its physical origin remains an active debate. Although several parameters seem to contribute to it {(e.g. particles size \citep{Guy_2015}, shape \citep{Brown_2011} or roughness \citep{Hsiao2016})}, it has become increasingly clear that frictional and non-contact interactions between particles play a key role \citep{Seto_2013,Mari_2015}. 
{Such} interactions 
{are} easily modified in numerical simulations, but {present} 
a real challenge 
in experiments. Consequently, 
{o}nly few experimental studies address
 the role of particle-particle interactions in dense suspensions rheology \citep[e.g.][]{Brown_2010,Lin_2015} {
{however lacking} systematic variation of these interactions.} Moreover, direct measurements of these interactions in relation to the rheology are also lacking so far.

{Here,}
we directly probe the microscopic interactions between individual particles and explore their link with the macroscopic rheology for dense cornstarch (CS) suspensions. The archetypical suspension {of}
CS grains in water exhibits a strong dynamic shear-thickening \citep{Merkt_2004,vonKann_2011,Waitukaitis_2012,Roche_2013}. Interestingly, Taylor \citep{Taylor_2013} shows that replacing water by polypropylene glycol in CS suspensions completely suppresses its shear-thickening nature and modifies its dielectric properties, 
reminiscent of observations in thermal suspensions \citep{Franks_2000,Cwalina_2014}.
Consequently, we tune particle interactions using suspending fluids with different chemical but similar physical properties (density, viscosity\dots). 
Specifically, we systemically study water/ethanol mixtures in different proportions combining three different techniques: \textit{(1)} non-steady-state rheology obtained from a sphere settling dynamics
; \textit{(2)} classical steady-state rheology; and \textit{(3)} atomic force microscopy (AFM) to probe particle interactions. 
{Both rheology techniques show that the typical shear-thickening behaviour observed for pure water turns into a low viscosity shear-thinning for pure ethanol, 
{passing} through {a} yield-stress-fluid state for intermediate mixtures. 
Furthermore, for water-based suspensions, shear-thickening and dynamic behaviours are observed, respectively in classical rheology and in settling experiments, at similar shear rates ranges.
We relate this to AFM measurements showing that particle interactions vary as the fluid is changed. Our results indicate that CS grains are covered by chemical agents behaving similarly to what was recently observed for polymer brushes \citep{Yu2015} exhibiting a cosolvency effect \cite{Cragg1946,Mukherji2016}. These dangling polymers may be at the origin of the peculiar rheology in water and also of the rheology changes with fluid as observed for colloidal suspensions \citep[e.g.][]{Mewis_2001,Wagner_2009}.

\textbf{\textit{Suspensions}} -- The suspensions are mixtures of CS particles in water-ethanol solutions. CS particles have irregular shape and diameters ranging from  $5$ to $20\ \mu m$. 
Freshly opened, 250~g sealed boxes of additive-free cooking CS were used. 
Density of CS from several boxes was determined by pycnometry: $\rho_\text{CS}\sim1542\pm15$~kg.m$^{-3}$.  The volume fraction of the suspensions is kept constant in this study: $\Phi^v_\text{CS}=40\%$. 
Although the {true} 
volume fraction 
might differ from $40\%$, due to, e.g., particle porosity and moisture contents \citep{Peters_2014}, our protocol ensures 
its reproducibility.
The suspending fluid consists of a mixture of demineralized water and ethanol (99.8\%) from Atlas \& Assink Chemie. We vary the mass fraction of the solution, $\Phi^m_\text{EtOH}$, from 0\% (pure water) to 100\% (pure ethanol). The suspending fluids are prepared one day before the experiment ensuring good mixing and cooling down. 

\textbf{\textit{Non-steady-state rheology}} -- The experimental set-up, 
shown in Fig.~\ref{Set-up_and_Measurement}a, 
consists of a cylindrical container (diameter $D=19.5$~cm, height $H=25$~cm) filled with the suspension into which we drop a sphere (mass $m_s=248$~g, radius $R_s=1.54$~cm). The release height $H_\text{fall}$ varies between -$2R_s$ (sphere starting immersed) and 30~cm. 
In order to follow the settling dynamics, a thin and rigid metal wire with tracers is attached to the top of the sphere. The mass of the wire ($\sim1$~g) and its resulting buoyancy force can be neglected compared to the sphere. We follow the tracers
 at a frame rate between 500 and 5000 Hz using a high speed camera (SA7, Photron).
{Correlating} successive images, we determine the sphere vertical position, $z$, velocity, $\dot{z}$, and acceleration, $\ddot{z}$, during its settling{.} 
\begin{figure}[t]
\includegraphics[scale=1]{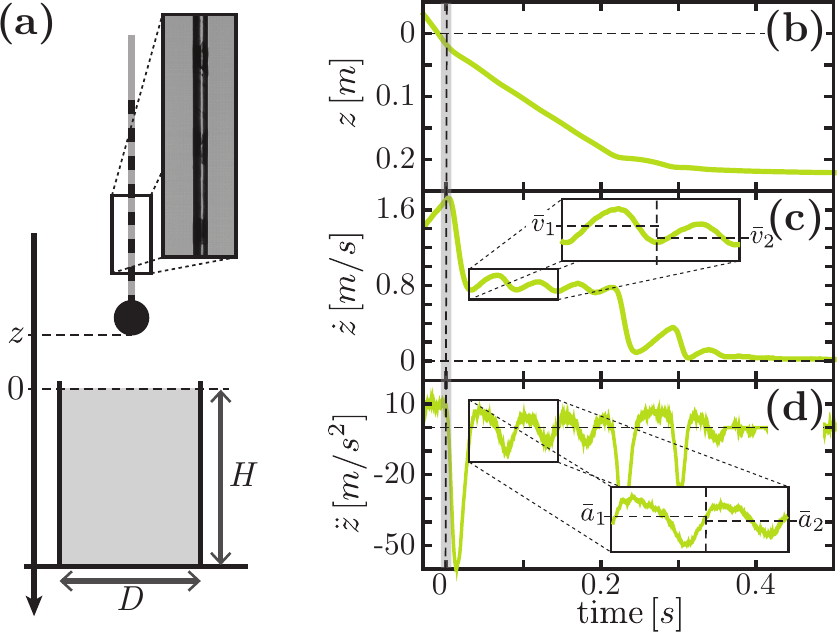}
\caption{$(a)$ Experimental set-up schematic. $(b-d)$ Typical time evolution {of} 
vertical position $z$, velocity  $\dot{z}$ and acceleration $\ddot{z}$ of the sphere for $\Phi^m_\text{EtOH}=$10\% and $H_\text{fall}=$15 cm. The insets in $(c)$ and $(d)$ show two zoomed-in oscillations with their respective mean velocities ($\bar{v}_1,\, \bar{v}_2$) and mean accelerations ($\bar{a}_1,\, \bar{a}_2$).}
\label{Set-up_and_Measurement}
\end{figure}

Fig.~\ref{Set-up_and_Measurement}b-d show the time evolution of $z$, $\dot{z}$ and $\ddot{z}$ for $\Phi^m_\text{EtOH}=$10\% and $H_\text{fall}=$15 cm. As previously observed for a CS suspension using pure water \citep{vonKann_2011,vonKann_2013}, after a rapid {slowing} down{ }
due to the impact{ }
(grey vertical line), $\dot{z}$ oscillates around a terminal velocity. For each oscillation we define its mean velocity, $\bar{v}_i$ and mean acceleration $\bar{a}_i$ (insets of Fig.~\ref{Set-up_and_Measurement}c and d). 
When approaching the bottom the sphere comes to a sudden full stop at $\sim$20 mm above the bottom. Then, it re-accelerates until it stops again. This repetitive stop-and-go behaviour is due to successive jamming and unjamming of the granular skeleton between the intruder and the bottom \citep{vonKann_2011}{.} 

When varying $\Phi^m_\text{EtOH}$, we observe a continuous change in settling dynamics (Fig.~\ref{Settling_Different_Phi}).
As $\Phi^m_\text{EtOH}$ is increased up to $\sim50\%$ the suspension viscosity rises (the average settling velocity decreases) and the oscillations disappear (Fig.~\ref{Settling_Different_Phi}a).
Beyond $\Phi^m_\text{EtOH} \sim$70\%, the viscosity becomes so small that the sphere bounces on the container bottom (Fig.~\ref{Settling_Different_Phi}b).

\begin{figure}[b!]
\includegraphics[scale=1]{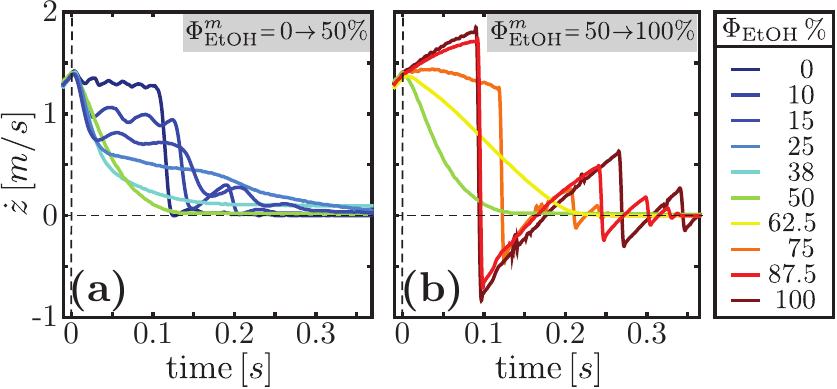}
\caption{Sphere velocity, $\dot{z}$, as a function of time for various $\Phi^m_\text{EtOH}$ and $H_\text{fall}=10$cm.}
\label{Settling_Different_Phi}
\end{figure}

\begin{figure*}[t!]
\includegraphics[scale=1]{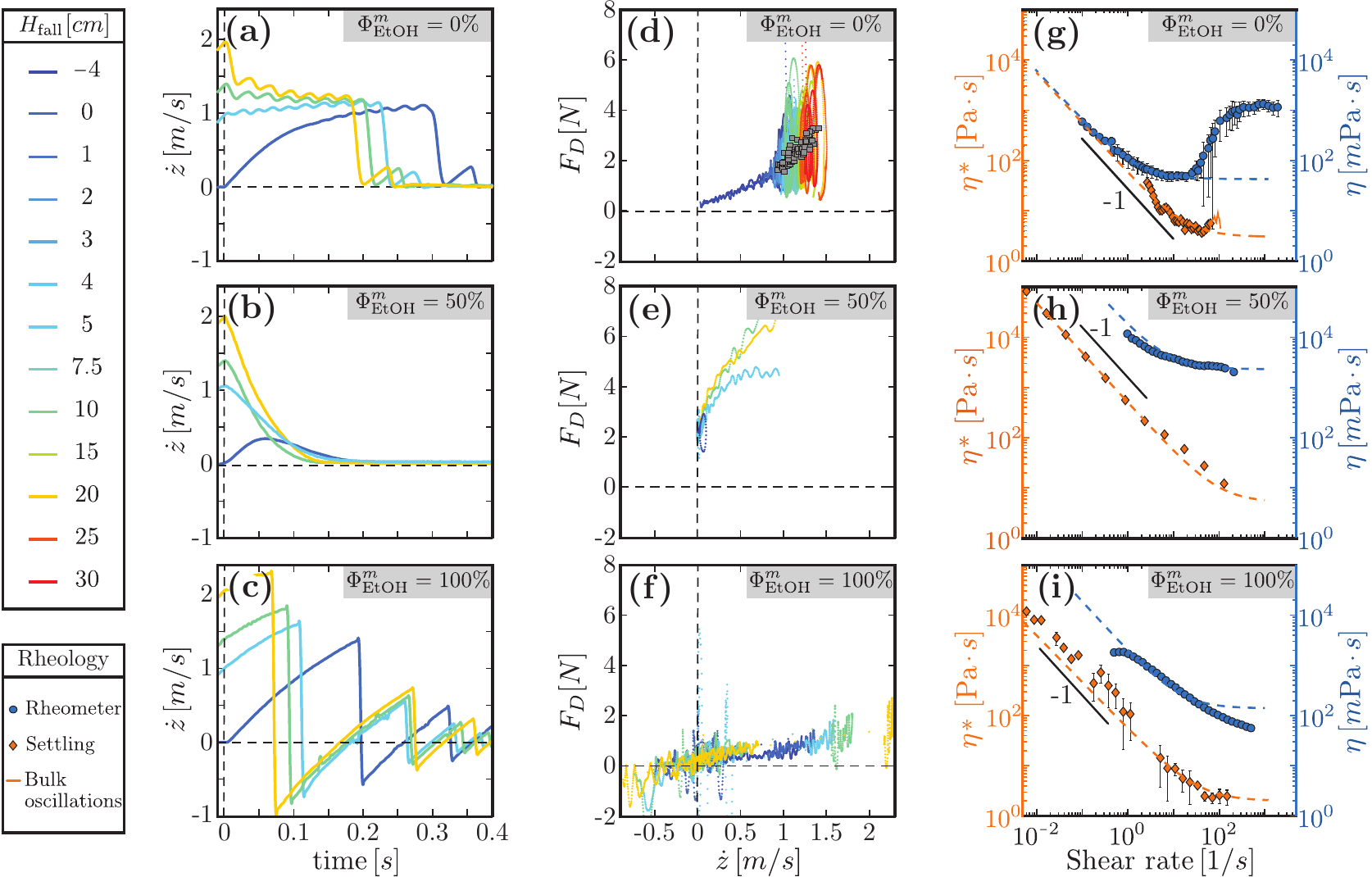}
\caption{\emph{Non-steady-state and classical rheology as a function of}  $\Phi^m_\text{EtOH}$: $(a$-$c)$ Sphere velocity, $\dot{z}$, as a function of time for various $H_\text{fall}$ 
$(d$-$f)$ Drag force, $F_D$, encountered by the sphere as a function of its velocity for the same and additional experiments. The grey squares in $(d)$ show the mean drag force as function of mean velocity $\bar{v}$ during oscillations. $(g$-$i)$ Flow curves from 
classical rheological measurements (blue circles) and apparent flow curves obtained from settling experiments (orange diamonds). {The dashed lines in $(g$-$i)$ are the best fits of the data with the Bingham model and} the orange line in $(g)$ corresponds to the bulk oscillations mean behaviour. }
\label{Settling_Data}
\end{figure*}

We can 
{distinguish} three typical behaviours{, }
illustrated in Fig.~\ref{Settling_Data}a-f. Panels a-c show the influence of the initial velocity for $\Phi^m_\text{EtOH}$=0, 50 and 100\% on the settling dynamics and panels d-f show the drag force, $F_D$, as a function of $\dot{z}$. 
{$F_D$} is derived from $\ddot{z}$ using the force balance on the sphere
\begin{equation}
F_D = m_s(\ddot{z}-g)+\frac{4}{3}\pi R_s^3\rho_{susp} g 
\end{equation}
in which $\rho_{susp}$ is the suspension density and $g$ is the gravitational constant.

{Up to $\Phi^m_\text{EtOH}\sim$20-25\%, the dynamics is similar to that of pure water }(Fig.~\ref{Settling_Data}a and \ref{Settling_Data}d).
Interestingly, the terminal velocity decreases with increasing $\Phi^m_\text{EtOH}$ but, 
is 
{independent} of the initial velocity, $V_0$. For low $V_0$, the sphere accelerates 
{towards} this terminal velocity. In contrast, for higher $V_0$, velocity decreases with oscillations reaching the same terminal velocity. $F_D$ increases linearly with $\dot{z}$ up to a critical velocity ($\sim$0.9m/s for pure water). Above this critical velocity, oscillations are observed and the 
{period averages $\bar{F}_D$} vs $\bar{v}$ (grey squares {in} Fig.~\ref{Settling_Data}d) collapse onto an unique curve whose slope seems to slightly increase with velocity{, corresponding} 
to an increased viscosity which is typical of a shear-thickening fluid.

For intermediate $\Phi^m_\text{EtOH}$ (between $\sim$20-25\% and $\sim$70\%), the sphere decelerates very rapidly after penetrating into the suspension. It then sinks at a constant velocity close to zero. 
{Consistenly,} Fig.~\ref{Settling_Data}e shows that $F_D$ equals the sphere weight in the limit of zero velocity. These two behaviours are typical of a yield-stress fluid when the density of the object is slightly above the critical density relatively to the yield stress \cite{Tabuteau_2007}. 

Finally, for $\Phi^m_\text{EtOH} \gtrsim$70\% (Fig.~\ref{Settling_Data}c and f), the sphere encounters a small drag resistance and bounces on the container bottom several times. Taking into account the noisy nature of the measurement,{ }
drag force curves for the different $H_\text{fall}$ collapse {onto} a single curve regardless of the velocity sign. 

\textbf{\textit{Classical rheology}} -- We use a MCR 502 rheometer (Anton Paar) with a concentric cylinders geometry. All measurements are repeated at least 3 times.
Fig.~\ref{Settling_Data}g-i presents the flow curves obtained from these rheological measurements (blue circles).  They are compared to the dynamical behaviour of the suspensions obtained from settling experiments (orange diamonds). To do such a comparison we define an apparent viscosity, $\eta*=F_D/(6\pi R_s\dot{z})$, and a characteristic shear rate, $\dot{\gamma}*=\dot{z}/R_s$. 
{Although Stokes' law is not applicable, it provide a reasonable estimation of the dynamic viscosity.}

The flow curves obtained from steady state classical rheology and from our dynamic system present a convincing qualitative agreement {although the numerical values are different{ }
probably due to {approximations} 
({Stokes' law) or }
geometrical factors}. For $\Phi^m_\text{EtOH}$ above $\sim$20-25\% these suspensions all present a yield stress and can be described by a simple Bingham equation : $\eta_B = \eta_{pl} + \sigma_Y/\dot{\gamma}$, in which  $\eta_{pl}$ is the plastic viscosity and $\sigma_Y$ the yield stress. This is consistent with 
{earlier observations in} CS suspensions {with} 
\{water/polypropylene glycol\} solutions \citep{Taylor_2013}.
For lower $\Phi^m_\text{EtOH}$, a Bingham equation can also approximate the flow curves for low shear rates. The values of $\sigma_Y$ as a function of  $\Phi^m_\text{EtOH}$ from both rheological measurements are shown {in} Fig.~\ref{Phase_Diagram}e. For both methods, $\sigma_Y$ reaches a maximum value for intermediate $\Phi^m_\text{EtOH}$ (between $\sim$25\% and $\sim$70\%). 
Finally, for low $\Phi^m_\text{EtOH}$ and high shear rates the steady-state rheology {exhibits} a strong shear-thickening which corresponds to the conditions in which bulk oscillations are observed during the settling experiments. 


\textbf{\textit{Particle-particle interactions}} -- We probe the interactions between CS particles using atomic force microscopy (AFM) by attaching  single CS grains to tipless cantilevers (see appendix for experimental details). We measure the force curves (Fig.~\ref{Phase_Diagram}a) while approaching and retracting this CS grain to{ }
other CS grains glued on the surface of a stainless steel disc in different water/EtOH solutions. 
From the force curves, we measure the adhesion force, $F_{adh}$, between individual CS grains 
and their apparent Young's modulus, $E^*${.} 
We also estimate an interaction length, $L_{int}${, 
corresponding} to the separation{ }
at which grains start to feel each other. Details on the analysis procedure can be found in the appendix.

Fig.~\ref{Phase_Diagram}a shows a typical force curve obtained in water{, representing} 
the{ }
force between the CS {particle} on the cantilever and one on the surface when approaching (blue) and retracting (red){.} 
On the retracting curve we observe sharp steps called pulling events. These events are signature{s} of high density dangling polymers 
{disentangling} in mediocre solvents \citep{Yu2015}.
This is a {plausible explanation} 
as CS is made of alternating semi-crystalline and amorphous layers of biopolymers amylose and amylopectin \citep{Buleon_1998}, being respectively slightly and mostly soluble in cold water \citep{Green_1975} but less and less soluble as ethanol is added to the solvent \citep[e.g.][]{Barrett1998,Zakrzewska2010} until being insoluble in ethanol \citep{GREENWOOD1970}.
Fig.~\ref{Phase_Diagram}b shows the percentage of measurements {with pulling events}
as a function of {$\Phi^m_\text{EtOH}$}. 
We observe them for all ethanol concentrations with a minimum for ethanol, logical with amylose and amylopectin solubilities, and 
a maximum for intermediate concentrations, which
indicates a cosolvency effect which is a solubility-maximum at intermediate $\Phi^m_\text{EtOH}$ \citep{Cragg1946,Yu2015,Mukherji2016}. 

\begin{figure}[t]
\includegraphics[scale=1.0]{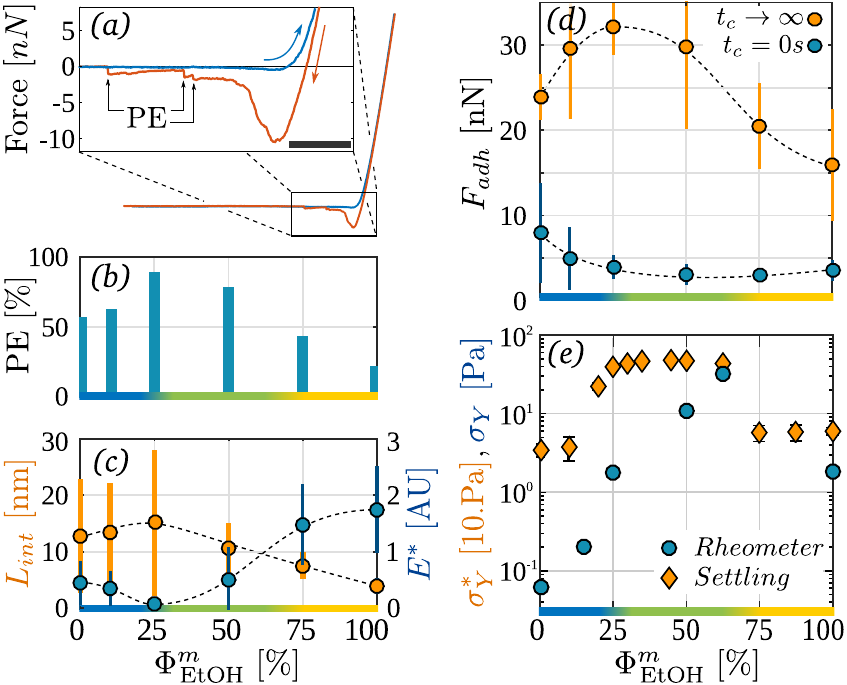}
\caption{\emph{Particle-particle interaction properties as a function of}  $\Phi^m_\text{EtOH}$: $(a)$ Typical force curve in water. Inset and arrows show the presence of pulling events (PE) while retracting (red curve). The black bar corresponds to a vertical displacement of 30~nm. $(b)$ Percentage of force curves exhibiting PE. $(c)$ Apparent Young's modulus, $E^*$ of an individual CS grain and interaction length, $L_{int}$, between CS grains. $(d)$ Adherence force, $F_{adh}$, between two CS grains for a contact time of $0s$ (blue) and in the limit of infinite contact time (orange). $(e)$ Yield stress measured from classical rheology (orange) and settling experiments (blue). The dashed lines are just guides to the eye. The color scales on the horizontal axis stand for the three typical behaviours: shear-thickening and bulks oscillation (blue), yield-stress fluid (green) and shear-thinning fluid (yellow).  } 
\label{Phase_Diagram}
\end{figure}

Fig.~\ref{Phase_Diagram}c shows {the} apparent elastic modulus {$E^*$ of one grain} (blue) and the interaction length, $L_{int}$, between two grains (orange). As the grains are not spherical and have sizes ranging from 5 to 20 $\mu$m, contact areas and curvature radii are difficult to assess, which is responsible for the large error bars. Therefore, we should not attach too much significance to the absolute values, but information from the data comparison for the different values of $\Phi^m_\text{EtOH}$ is to be trusted. 
Thus we observe that the apparent {particle} softness 
and the {interaction} length 
{vary} 
{with} $\Phi^m_\text{EtOH}$ which we interpret as a result of the cosolvency effect{:} 
$\Phi^m_\text{EtOH}\approx 25\%$ {appears to be 
the best solvent{, which is consistent with more pulling events being} 
observed for this concentration. Indeed, a better solvent allows for deeper interdigitation of the polymers in opposing grains. {Although $L_{int}$ varies inversely to $E^*$, from our data it is not possible to determine the origin of the repulsive force before elastic contact: it could be either interdigitation of the polymers or some form of non-contact repulsion, like static charges.}

After the approach, it is possible to keep {two CS} grains in contact for a given contact time {$t_c$ before retracting.} Doing so we can investigate the effect of the contact time on the adherence force {$F_{adh}$}{.} 
Fig.~\ref{Phase_Diagram}d shows the evolution of {$F_{adh}$} as a function of  $\Phi^m_\text{EtOH}$ for {zero} contact time, $F_{adh}^0$ (blue), and in the limit of infinite contact time, $F_{adh}^\infty$ (orange). For all $\Phi^m_\text{EtOH}$, $F_{adh}$ increases with $t_c$ following an exponential {decay} 
characterized by a {decay} time $\tau$ (see appendix for details){, which is again consistent with our interpretation of free dangling polymers interpenetrating with time.}  
We observe that $F_{adh}^0$ is maximal in pure water whereas $F_{adh}^\infty$  exhibits a maximum for $\Phi^m_\text{EtOH} = 25\%$. We attribute the latter to the larger effective interaction area due to particle softness and polymer interpenetration.

These results are consistent with the macroscopic rheology. Indeed, $F_{adh}^\infty$ must be related to the suspension behaviour at small shear rate, i.e., the yield stress $\sigma_Y$, which we obtain from the Bingham fits to the flow curves of Fig.~\ref{Settling_Data}. Although slightly shifted, $F_{adh}^\infty$ shows similar variations as the yield stress extracted from rheology experiments (Fig.~\ref{Phase_Diagram}e). Moreover, the shear-thinning part of the flow curves observed for all $\Phi^m_\text{EtOH}$ is also consistent with an increase of $F_{adh}$ with $t_c$. 
On the other hand, one could expect that the suspension behaviour at high shear rate could be related to $F_{adh}^0$ and $\tau$. But present measurements don't show any quantitative indication of that, although $\tau$ does vary with $\Phi^m_\text{EtOH}$. Namely, $\tau$ is minimal for pure water ($\tau_{min}=0.5\pm0.1s$) and maximal for intermediate concentration ($\tau_{min}=2.0\pm0.7s$) (see appendix). 
Therefore, friction measurements as described in \citep{Fernandez_2015} could provide additional insights \citep{Fernandez_2013,Hsiao2016}, although with interpretation difficulties due to CS particle irregularity and  roughness.

\textbf{\textit{Summary}} -- In this letter we show that {gradually} replacing the suspending fluid of the well-known suspension of cornstarch and water by ethanol, the familiar shear-thickening behaviour completely disappears. Going from pure water to pure ethanol, the suspension behaviour changes continuously with ethanol concentration from dynamic shear-thickening for pure water to low viscosity shear-thinning  for pure ethanol, passing through a yield-stress fluid phase for intermediate mixtures.
Comparison of classical (steady-state) and non-steady-state settling rheology shows qualitative agreement. More specifically, it shows that flow conditions for which shear-thickening is observed in classical rheology measurements correspond to the conditions for which bulk oscillations are observed in non-steady-state experiments.

These behaviours are related to the interactions between CS grains in the different suspending fluids {measured} using atomic force microscopy. We first present evidence that CS grains are covered  by free dangling polymers behaving like polymer brushes. Then, the variation of the adherence force with the suspending fluid is shown to be consistent with the yield stress observed in macroscopic rheology. This indicates that the macroscopic behaviour is closely linked to the details of the particle-particle interactions.
It appears that the presence of dangling polymers may not only be at the origin of the strikingly different behaviours observed while changing the suspending fluid but also of the peculiar dynamic behaviour of suspensions of CS in water. 
In order to validate this hypothesis, it is essential to perform additional research on better controlled systems such as suspensions of spherical particles functionalized with known polymer brushes.


--{\it Acknowledgments.} The authors thank H. Gojzewski for preparing the cantilevers. This work is part of a research programme of the Foundation for Fundamental Research on Matter (FOM), which is financially supported by the Netherlands Organisation for Scientific Research (NWO). LOG is 
supported by Becas Chile - CONICYT and AP by a FOM/v post-doctoral grant.

\appendix*
\section{Atomic Force Microscopy (AFM)}

\subsection{Experimental Details}
We use a Bruker AFM (Multimode 8 with a Nanoscope V controller) using a JV vertical engage scanner and a Bruker glass liquid cell. 
Using a micromanipulator and UV curing glue (NOA 81) we attached a single cornstarch particle to the end of three tipless AFM cantilevers (TL-CONT-50, sQube, Germany){, with spring constants} 
$3.24$, $2.93$ and $2.61\ \ N.m^{-1}$ and 
resonance {frequencies} of $85$, $75$ and $79\ kHz$ respectively (Fig.~\ref{SEM}a). 
The tested surfaces consist of  stainless steal discs covered with cornstarch particles{,   
attached} using {an} epoxy two-component glue (Fig.~\ref{SEM}b). 
{F}orce curves are measured while approaching and retracting the cornstarch grain to and from these surfaces in different water/EtOH solutions ($\Phi^m_\text{EtOH}=$~$ 0,\ 10,\ 25,\ 50,\ 75,$ and $100\%$) with a velocity of 0.77, 1.44 to 2.88~$\mu m.s^{-1}${, for which the }
analysis shows no influence of the approaching and retracting velocity. 

\begin{figure}[h!]
\includegraphics[scale=0.95]{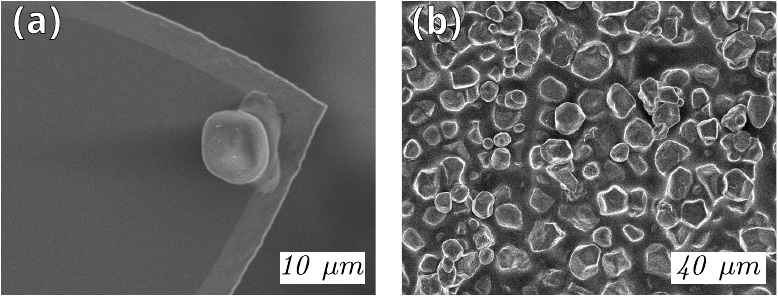}
\caption{SEM image of $(a)$ a cornstarch grain glued to the end of a tipless AFM cantilever and $(b)$ a sample surface covered {with} 
cornstarch grains.}
\label{SEM}
\end{figure}

With the different cantilevers, we probe 3 positions on 2 different samples and compared them to reference force curves measurements on bare glue to ensure that we truly probe the CS-CS interactions. For each position, force curves are averaged over at least 50 measurements. Fig.~\ref{Phase_Diagram} shows 
averaged results for one cantilever/sample set.
{O}ther cantilevers{/sample} show {similar} 
trends. 

\subsection{Force Curve Analysis}

The force measured upon close approach {was}
fitted with the Hertzian contact model following the procedure presented {in }
\citep{Lin_2007a,Lin_2007b}
{to} obtain $E^*$ (Fig.~\ref{Phase_Diagram}c) and{ }
the position at which the contact starts to become elastic. This point is considered as the contact point.
As{ }
the cornstarch grains are not perfectly round and their radius not well defined, we only obtain an apparent modulus. Thus,{ }
absolute values 
have no concrete{ }
interpretation but can be compared for{ }
different suspending fluids. The maximum force reached is $60-80\ nN$. 

The adhesion $F_{adh}$ (Fig.~\ref{Phase_Diagram}d) is the force 
just before the cornstarch grains snap out of contact while retracting the cantilever.

\subsection{Effect of Contact Duration on the Adhesion}

As the rheology of cornstarch suspensions is observed to strongly depend on the shear rate we study the evolution of the adhesion force as a function of the contact duration between two  grains. 
{To do so, we approach the grain to the surface, keep} grains in contact during a waiting time, $\Delta t$, ranging from 0 to 20 s{, and then retract.} These measurements are performed in different water/EtOH solutions with a velocity of 1.44~$\mu m.s^{-1}$. 

As the geometry of the contact may vary from one probing position to another, for each position we normalize the adhesion force by the one corresponding to zero waiting time:
\begin{equation}
\tilde{F}_{adh}(\Delta t) = \frac{F_{adh}(\Delta t)}{F_{adh}(\Delta t=0s)}
\end{equation}
 Figure~\ref{effect_WT}a shows the variation of the normalized adhesion force $\tilde{F}_{adh}$ as a function of the waiting time for $\Phi^m_\text{EtOH}= 0\%$ and for each different probing position. The normalized adhesion force can be fitted by an exponential 
\begin{equation}
\tilde{F}_{adh}(\Delta t) = \tilde{F}_{adh}^\infty - (\tilde{F}_{adh}^\infty -1 )\exp(-\frac{\Delta t}{\tau})
\end{equation} 
$\tilde{F}_{adh}^\infty $ and $\tau$ are measured for each position and for each $\Phi^m_\text{EtOH}$. The values shown in Fig.~\ref{Phase_Diagram}d and Figure~\ref{effect_WT}b correspond to the average over all positions for each  $\Phi^m_\text{EtOH}$.

\begin{figure}[h!]
\includegraphics[scale=1]{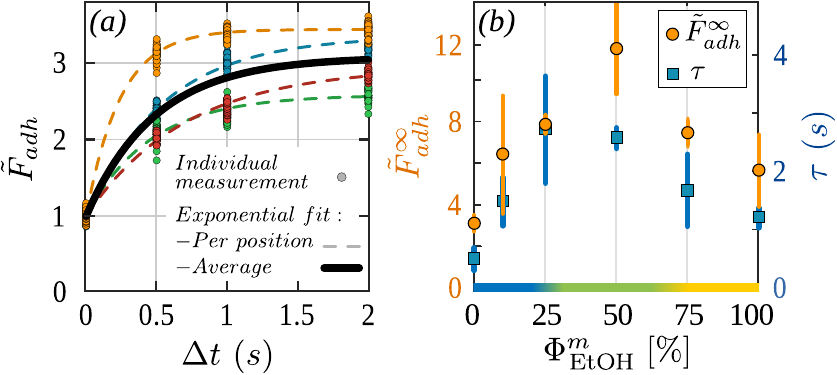}
\caption{$(a)$ Variation of $\tilde{F}_{adh}$  as a function of waiting time. Colors stand for each probing position. Dashed lines correspond to the exponential fit for each position and the solid line is the average fit. $(b)$ Average $\tilde{F}_{adh}^\infty $ and $\tau$ as a function of  $\Phi^m_\text{EtOH}$ }
\label{effect_WT}
\end{figure}


%

\end{document}